\documentclass{article}

\usepackage{tikz}           
\usetikzlibrary{arrows.meta, positioning} 
\usepackage{arxiv}          
\usepackage[utf8]{inputenc} 
\usepackage[T1]{fontenc}    
\usepackage{hyperref}       
\usepackage{url}            
\usepackage{booktabs}       
\usepackage{amsfonts}       
\usepackage{nicefrac}       
\usepackage{microtype}      
\usepackage{natbib}         
\usepackage{doi}            
\usepackage{caption}        

\tikzstyle{mainblock} = [rectangle, draw, fill=blue!20, text width=6em, text centered, rounded corners, minimum height=4em]
\tikzstyle{subblock} = [rectangle, draw, fill=blue!10, text width=5em, text centered, rounded corners, minimum height=3em]
\tikzstyle{line} = [draw, -Latex]

\title{Music Information Retrieval on Representative Mexican Folk Vocal Melodies Through MIDI Feature Extraction}

\author{ \href{https://orcid.org/0009-0009-3230-8476}{\includegraphics[scale=0.06]{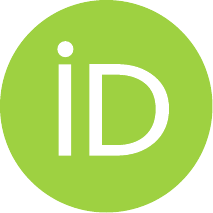}\hspace{1mm}Mario Alberto Vallejo Reyes} \\
School of Physics, Engineering and Technology\\
University of York\\
York, United Kingdom \\
\texttt{mario.vallejo@york.ac.uk} \\
}

\date{}

\renewcommand{\headeright}{}
\renewcommand{\undertitle}{}
\renewcommand{\shorttitle}{Music Information Retrieval on Representative Mexican Folk Vocal Melodies Through MIDI Feature Extraction}

\begin{document}
\maketitle

\begin{abstract}
This study analyzes representative Mexican folk vocal melodies using MIDI feature extraction, examining ambitus, pitch-class entropy, and interval distribution.  It also explores the relationship between these features and song popularity, as measured by Spotify plays. The study employs MATLAB and the MIDI Toolbox for extracting musical features and performing statistical analysis. The findings reveal a significant variation in ambitus, with values ranging from 8 to 27 semitones, indicating a diverse compositional style and vocal demand across the genre. The analysis of pitch-class entropy showcases a broad spectrum of melodic complexity, with Armando Manzanero's `Somos Novios' displaying the highest entropy, suggesting varied and complex melodic structures, while traditional pieces like `La Bamba' exhibit lower entropy, indicating simpler, more repetitive patterns. The interval distribution predominantly features prime intervals (P1), major and minor seconds (M2, m2), pointing to a compositional preference for close, contiguous intervals that contribute to the melodies' accessibility and appeal. Statistical analysis do not establish a significant correlation between the ambitus or entropy and the number of Spotify plays.
\end{abstract}


\section{Introduction and Background}

\subsection{Research Objectives and Significance}
The primary objective of this study is to provide a comprehensive analysis of Mexican vocal melodies through MIDI feature extraction, focusing on ambitus, pitch-class entropy and interval distribution. The research focuses on critical questions designed to explore the intrinsic qualities of Mexican folk melodies: How do the variations in pitch range, as measured by ambitus, reflect the diversity of melodic expression in Mexican folk music? What insights can be gained about the complexity of these songs through the analysis of pitch-class entropy? And lastly, what are the discernible patterns in interval distribution within Mexican folk music, and how do they contribute to its melodic essence?

Another exploratory goal of this research is to probe whether there is a correlation between the musical characteristics of entropy and ambitus and the song's popularity, as measured by Spotify play counts. This part of the study explores new perspectives in MIDI feature extraction, particularly the intersection between technical musical analysis and audience engagement in a digital era.

These objectives are to contribute to the broader knowledge of musicology, computational music analysis, and digital musicology, providing insights and data that may be useful for future research and practical applications in these fields.

\subsection{Review of Recent Methods}
A detailed search was conducted in various academic databases, including Google Scholar, to review recent methodologies in the extraction of MIDI features. This search specifically focused on studies in the Music Information Retrieval (MIR) field, emphasising MIDI feature extraction in folk music. Through this process, Anna Maria Christodoulou's thesis, `Computational Analysis of Greek Folk Music of the Aegean Islands'\citeyearpar{3228994}, emerged as a study of particular relevance.

The analysis of Christodoulou's study was based on its direct relevance to MIDI feature extraction in folk music and its comprehensive methodological approach. Examining Christodoulou's approach offered another perspective before researching mexican vocal melodies.

In her thesis, Christodoulou utilised a methodical approach to music analysis, beginning with selecting Greek folk music from the Aegean islands. This selection was pertinent to ensure a proper scope for analysis, enabling a concentrated examination of the musical elements unique to this region's folk music.

The primary methodological step in Christodoulou's research involved extracting musical features from MIDI and WAV files, employing tools such as the MIDI Toolbox. This phase is critical in computational music analysis, as it allows for systematic identification and quantification of various musical elements, including pitch, rhythm, and melody patterns. The extraction laid the groundwork for the subsequent pattern discovery phase.

In this phase, Christodoulou focused on identifying recurring motifs and structural patterns within the selected music. The objective was to uncover the underlying patterns and structures of Aegean island folk music.

Christodoulou also applied Machine Learning techniques, such as Self-Organizing Maps and t-distributed Stochastic Neighbor Embedding, for the unsupervised clustering of the extracted features. This method facilitated the systematic grouping of similar musical elements, contributing to a more organised analysis and revealing broader musical relationships and trends.

Finally,  an essential aspect of her methodology included a comparative analysis with existing studies in the field. By situating her findings within the larger scope of musicological research, Christodoulou added context to her approach and contributed new perspectives on Greek folk music compared to other regional music traditions.

\subsection{Succinct Introduction to the Chosen Approach}
This study utilised a MATLAB-based pipeline to analyse MIDI vocal melodies representative of Mexican folk music, incorporating the MIDI Toolbox for data processing and analysis. MIDI files, normalised using Pro Tools, were imported into MATLAB. The MIDI Toolbox was used to extract features such as ambitus, pitch-class entropy, and interval distribution from these files.

In addition to feature extraction, MATLAB was employed to generate plots and conduct statistical analyses. This approach facilitated a streamlined workflow, allowing for the visual representation of data and analysis of the extracted features. The use of MATLAB's statistical tools aided in examining patterns and trends within the dataset.

Overall, this methodology enabled the examination of the characteristics of Mexican folk music as represented in the MIDI vocal melodies, combining the functionalities of MATLAB and the MIDI Toolbox.

\section{Methodology and Implementation}

\subsection{MIR Pipeline Design}
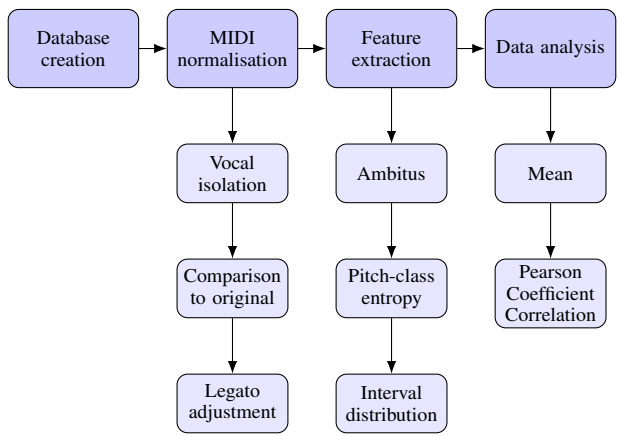
\begin{figure}[htbp]
\centering
\begin{tikzpicture}[node distance=1cm and 0.5cm, auto, scale=0.74, every node/.style={transform shape}]
    \tikzstyle{mainblock} = [rectangle, draw, fill=blue!20, text width=6em, text centered, rounded corners, minimum height=4em]
    \tikzstyle{subblock} = [rectangle, draw, fill=blue!10, text width=5em, text centered, rounded corners, minimum height=3em]

    \node [mainblock] (dbcreation) {Database creation};
    \node [mainblock, right=of dbcreation] (midinorm) {MIDI normalisation};
    \node [mainblock, right=of midinorm] (featureextraction) {Feature extraction};
    \node [subblock, below=of featureextraction] (ambitus) {Ambitus}; 
    \node [subblock, below=of ambitus] (entropy) {Pitch-class entropy}; 
    \node [subblock, below=of entropy] (intervaldistribution) {Interval distribution};
    \node [mainblock, right=of featureextraction] (dataanalysis) {Data analysis};

    \node [subblock, below=of midinorm] (vocalisolation) {Vocal isolation};
    \node [subblock, below=of vocalisolation] (comptooriginal) {Comparison to original};
    \node [subblock, below=of comptooriginal] (legatoadjustment) {Legato adjustment};

    \node [subblock, below=of dataanalysis] (mean) {Mean};
    \node [subblock, below=of mean] (pearson) {Pearson Coefficient Correlation};

    \path [line] (dbcreation) -- (midinorm);
    \path [line] (midinorm) -- (featureextraction);
    \path [line] (featureextraction) -- (dataanalysis);
    \path [line] (featureextraction) -- (ambitus);
    \path [line] (ambitus) -- (entropy);
    \path [line] (entropy) -- (intervaldistribution);

    \path [line] (midinorm) -- (vocalisolation);
    \path [line] (vocalisolation) -- (comptooriginal);
    \path [line] (comptooriginal) -- (legatoadjustment);

    \path [line] (dataanalysis) -- (mean);
    \path [line] (mean) -- (pearson);
\end{tikzpicture}
\caption{Schematic Overview of the Music Information Retrieval Process}
\end{figure}

This study's Music Information Retrieval (MIR) pipeline was structured to analyze MIDI vocal melodies of Mexican folk music. The study's design included the development of a pipeline for efficient data processing and analysis, which comprised the following stages: The initial phase of the study involved the meticulous selection of MIDI files for Mexican folk songs that are emblematic of the genre. Each song was chosen for its cultural significance and its impact on the tapestry of Mexican music. Factors such as the song's historical importance, its role in cultural and national events, its presence in popular media, and its recognition by prestigious institutions and in global music charts were considered. This approach ensured a diverse and culturally rich sample representing Mexican folk music's essence. The MIDI files, sourced from online databases, were chosen with an emphasis on their accuracy and fidelity to the original melodies.

After collection, the MIDI files underwent preliminary processing in the digital audio workstation Pro Tools to prepare them for the feature extraction phase of the analysis. This preparatory stage was instrumental in ensuring the MIDI files were in the appropriate format to facilitate the subsequent extraction of musical features. 

The core of the MIR pipeline, this phase involved extracting specific musical features from each MIDI file using MATLAB. The features extracted included ambitus, pitch-class entropy and interval distribution. This selection was guided by their relevance to a comprehensive understanding of Mexican folk music's musical structure and characteristics. 

The pipeline's final stage focused on analysing and interpreting the extracted features. This study stage involved employing various analytical tools and methods to synthesize the data and identify patterns and trends within the melodies. This included using graphical methods and statistical analysis to interpret the findings and draw conclusions about the musical elements that define Mexican folk music.

\subsection{Feature Extraction and Normalisation}
The project utilised MATLAB for a detailed feature extraction and normalisation process, focusing on ambitus, pitch-class entropy, and interval distribution in vocal melodies. This process included the integration of automated text file generation and data visualisation within the MATLAB code.

The ambitus calculation was performed using the MIDI Toolbox's \texttt{ambitus} function. The MATLAB script began by identifying the MIDI file folder and iterating through each file to read it into a note matrix (nmat). For each MIDI file, the script used the \texttt{ambitus} function to calculate the musical range in semitones. The script was programmed to create and write a text file to document these calculations, ensuring systematic recording of the ambitus data.

Using MATLAB's plotting functions, the script generated a bar graph to represent the ambitus values visually. This graph was a direct output from the analysis process, visually comparing the musical ranges across the dataset.

Entropy analysis involved processing each MIDI file to calculate the pitch class distribution, subsequently feeding this data into a custom \texttt{entropy} function. This function calculated the entropy value by normalising the pitch class distribution and applying logarithmic calculations. The script automatically recorded the results in a text file and generated a bar plot to compare entropy values across different melodies visually.

Finally, the interval distribution analysis used the \texttt{ivdist1} function to calculate the frequency of melodic intervals for each song. The script aggregated and normalized these distributions and automatically generated a plot to display the normalized aggregate interval distribution, offering a comprehensive view of common interval patterns.

\subsection{Selection of Features}
The MIDI Toolbox offers a variety of functions for extracting musical features from MIDI files. This study selectively focused on features aligning with the analytical goals of Mexican vocal melodies. The decision was influenced by the fact that many functions in the MIDI Toolbox, like pitch distribution, are key-specific and less relevant for this study. Considering Mexican songs are often performed in various keys, depending on the singer's preference, a key-specific analysis is less pertinent for a holistic understanding of the genre.

Therefore, the features selected for this study — ambitus, pitch-class entropy, and interval distribution — are critical for analyzing melodic aspects that are not dependent on key variations. This approach enables a more objective analysis of the musical elements consistent across different key interpretations within the genre.

\subsubsection{Ambitus}
The ambitus function quantifies the pitch range in a given MIDI file, ascertained by the difference between the highest and lowest notes, and serves as an objective metric for analyzing melodic scope. In analyzing folk Mexican vocal melodies, this measurement is vital for identifying the range of pitches chosen by composers and the ensuing vocal demands placed on performers. This dual perspective enhances our understanding of the technical complexity and range required in Mexican music, contributing to a deeper appreciation of its vocal traditions.

\subsubsection{Pitch-class Entropy}
Pitch-class entropy is a statistical measure used to quantify the diversity in pitch distribution of a musical composition. This measure is derived by calculating the Shannon entropy of the pitch class distribution, yielding a numerical value that reflects the level of predictability or unpredictability in pitch usage. High entropy values indicate compositions with varied and complex pitch patterns, whereas low values suggest more straightforward, repetitive pitch sequences. 

In the study of Mexican vocal melodies, pitch-class entropy is applied to assess the complexity of the compositions objectively. Through this method, it can be determined whether these melodies are characterized by a broad range of pitch variations or by more consistent and uniform pitch patterns.

\subsubsection{Interval Distribution}
The interval distribution function, as described in the MIDI Toolbox manual by Eerola et al.\citeyearpar{Eerola2004}, calculates the frequencies of various musical intervals from a sequence of notes. It presents these frequencies in terms of semitones, providing a clear profile of interval usage within a piece. 

For the study of Mexican vocal melodies, interval distribution is a crucial tool for examining the patterns of interval usage in these compositions. Understanding the prevalence and variety of intervals provides a clearer comprehension of the compositional characteristics that define Mexican vocal music.

\subsection{Statistical Analysis Methods}
This study employed descriptive and inferential statistical methods to analyze features extracted from Mexican vocal melodies. The initial phase involved calculating the mean for the ambitus and entropy of the melodies, which helped establish the central tendencies within the dataset.

The study then utilized inferential statistics, particularly the Pearson correlation coefficient, to investigate the relationship between musical features—ambitus and entropy—and song popularity, as reflected by Spotify play counts. The Pearson correlation coefficient was used to measure the strength and direction of the association between these musical features and song popularity.

Further, the significance of these correlations was assessed using p-values. This step was crucial for determining whether the correlations were statistically significant or could have occurred by chance, thereby evaluating musical features' potential influence on the songs' popularity within the digital landscape.

\section{Experimentation and Results}

\subsection{Dataset Creation}
The compilation of representative Mexican songs in this study is characterised by their widespread recognition and cultural significance. These songs have been selected based on thorough research to establish their representative status. The subsequent sections provide an in-depth analysis of each song, presenting factual evidence of their cultural importance and relevance within Mexican society.

`Cielito Lindo' is a unifying anthem at international sports events, especially during Mexico's World Cup matches. It has also resonated in times of national crisis, notably being sung in the aftermath of the catastrophic earthquake in Mexico in 2017. This song is integral to cultural and familial celebrations, from weddings to national holidays. It has been popularised globally with renditions by Plácido Domingo, a world-renowned operatic tenor with multiple Grammy Awards, and Luciano Pavarotti, one of the most celebrated operatic voices of the 20th century\cite{CielitoLindoWP}.

`La Bamba' is recognised for its significant cultural impact, as the US Library of Congress has documented over 1,000 verses variations of the song by different authors. It was honoured in the Latin Grammy Hall of Fame and is the only Spanish song listed in Rolling Stone's 500 Greatest Songs of All Time. The 1987 cover by Los Lobos reached number one internationally. Notably, it was used as a counter-anthem to a white supremacist rally, symbolising resistance against divisive ideologies\citeyearpar{LaBambaFT}.

`México Lindo y Querido' holds a special place in Mexican culture, being integral to events such as Independence Day celebrations, weddings, and quinceañeras\citeyearpar{MexicoLindoSongMeaning}. It was also prominently featured in movies during the Golden Era of Mexican cinema from the 1940s to `60s\citeyearpar{RancheraNPR}. In a symbolic act of unity and national pride, it was performed by a children's choir on the US-Mexico border in response to Donald Trump's proposed border wall\citeyearpar{MexicoChoirBorder2017}. The funeral of Jorge Negrete, the artist who made the song famous, was attended by half a million people\citeyearpar{NegreteFuneral2020}.

`Guadalajara' has achieved international acclaim and has been performed by artists like Elvis Presley, demonstrating its global recognition and appeal. `Guadalajara' functions as a cultural ambassador, mentioning in its lyrics Mexican cities, traditions, and gastronomy, with its composer, Pepe Guízar, earning the nickname `The musical painter of Mexico'\citeyearpar{guadalajaramilenio}. In the context of national patriotic celebrations and traditional mariachi performances, it is a prevailing occurrence to hear the interpretation of `Guadalajara'\citeyearpar{mexicodesconocido}.

`La Cucaracha' is a song that has adapted to various historical periods in Mexico, such as the Mexican Revolution, with its lyrics being modified to reflect the sentiments of each period\citeyearpar{cucaracha_shadow}. The US Library of Congress preserves a 1915 broadside of `La Cucaracha', emphasising its historical and cultural importance\citeyearpar{lacucaracha1915}. The song has been featured in Looney Tunes cartoons and inspired a 1934 musical short film that won an Oscar for Best Comedy Short Film\citeyearpar{nuestrostories2022cucaracha}.

`El Son de la Negra' is considered by many as Mexico's second national anthem\citeyearpar{lamusart_mariachi}\citeyearpar{byu_marriott}\citeyearpar{nytimes_mariachi}. It was included in the ``Collection of Musical Testimonies'' by Mexico's National Institute of Anthropology and History. It gained international acclaim when featured in the Museum of Modern Art in New York's exhibition ``Twenty Centuries of Mexican Art''\citeyearpar{economista-son-negra}. `El Son de la Negra' has two main vocal melodies, which sing to the same rhythm but have different melodies. In this study, both melodies were analysed and are referred to as `El Son de la Negra High Lead' and `El Son de la Negra Low Lead'.

`Sabor a Mí' is recognised, particularly for the version by Javier Solis, which was inducted into the Latin Grammy Hall of Fame\citeyearpar{frontera-sabor-a-mi}. Sabor a Mí has been covered by internationally acclaimed artists, including Eydie Gormé, the K-pop band Exo\citeyearpar{milenio-sabor-a-mi}, Yoshiro Hiroishi\citeyearpar{muyinteresante-sabor-a-mi}, Kenny G and Placido Domingo, among others, underscoring its worldwide recognition.

`El Rey' has been included in both film and television productions, such as ``Breaking Bad'' and ``Narcos''; this indicates its influence on popular culture, both domestically in Mexico and on a global scale. Furthermore, `El Rey' has become a staple in Mexican karaoke bars and fiestas, emphasising its enduring popularity and cultural significance in social gatherings\citeyearpar{meaning-el-rey}. Ranked \#19 in Billboard's list ``Best Latin Songs of All Time''\citeyearpar{billboard}.

`Bésame Mucho' has been interpreted by some of the most famous artists of all time, including The Beatles, Frank Sinatra, and Elvis Presley\citeyearpar{BesameMuchoGuardian2005}. Andy Russell's cover became an anthem for lovers separated by World War II\citeyearpar{BesameMuchoNuestroStories}. Statues commemorating the song's author have been erected in several Mexican cities, including Mexico City\citeyearpar{ConsueloVelasquezBesameMucho}, Ciudad Guzmán\citeyearpar{ZapotlanGovArticle}, and Guadalajara\citeyearpar{RotundaNewIllustriousCitizens}.

In its original Spanish version, `Somos Novios' remains one of the most covered Spanish-language songs of all time. The English adaptation of the song, titled `It's Impossible', received a Grammy nomination for Song of the Year in 1971 and reached the top 10 of the Billboard Hot 100. Additionally, this version was covered by various highly renowned international artists, including Elvis Presley, Andy Williams, Shirley Bassey, and Andrea Bocelli, among others, further solidifying its status as a globally celebrated and representative Mexican song\citeyearpar{billboard-manzanero-songs}.

\subsection{Dataset Pre-processing}
The dataset employed for this study underwent a systematic process to ensure data accuracy and suitability for the anal- ysis of ambitus, pitch-class entropy, and interval distribution. Initially, MIDI files corresponding to the selected songs were gathered from online sources. These files were then imported into Pro Tools, a Digital Audio Workstation, for further refinement.
Since MIDI files typically include multiple instrument tracks, the primary goal was to isolate the vocal melodies. To verify the accuracy of the extracted vocal melodies, a meticulous comparison with the original song recordings was conducted. Additionally, legatos within the vocal melodies were adjusted to prevent note overlap, as MIDI Toolbox analysis requires non-overlapping monophonic data.
Following these necessary pre-processing steps, the modified MIDI files, now containing isolated and non-overlapping vocal melodies, were exported. These processed MIDI files were integrated into the dataset, ensuring that the dataset accurately represented the vocal melodies of the selected songs. The procedures undertaken during dataset preparation guarantee the reliability of subsequent research findings.

\subsection{Presentation of Results}
\subsubsection{Ambitus}
\begin{figure}[htbp]
    \centering
    \begin{minipage}{0.48\textwidth}
        \centering
        \includegraphics[width=\linewidth]{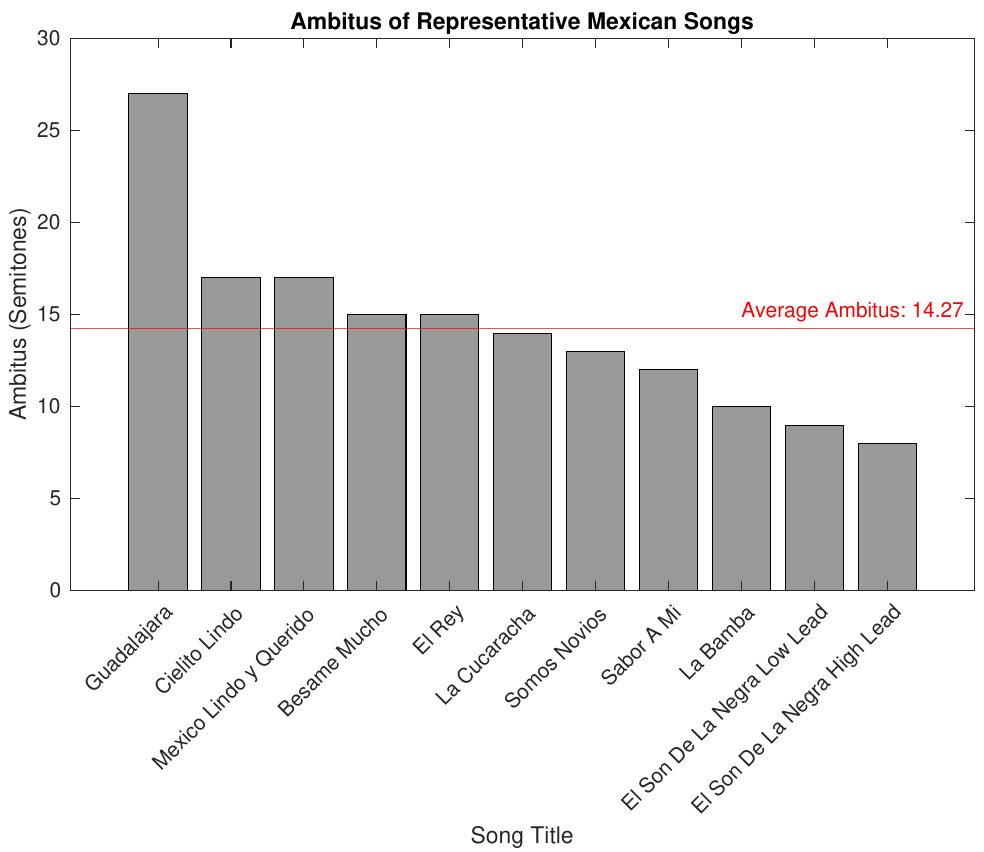}
        \caption{Average Ambitus of Selected Mexican Vocal Melodies}
        \label{fig:ambitus1}
    \end{minipage}
    \hfill
    \begin{minipage}{0.48\textwidth}
        \centering
        \includegraphics[width=\linewidth]{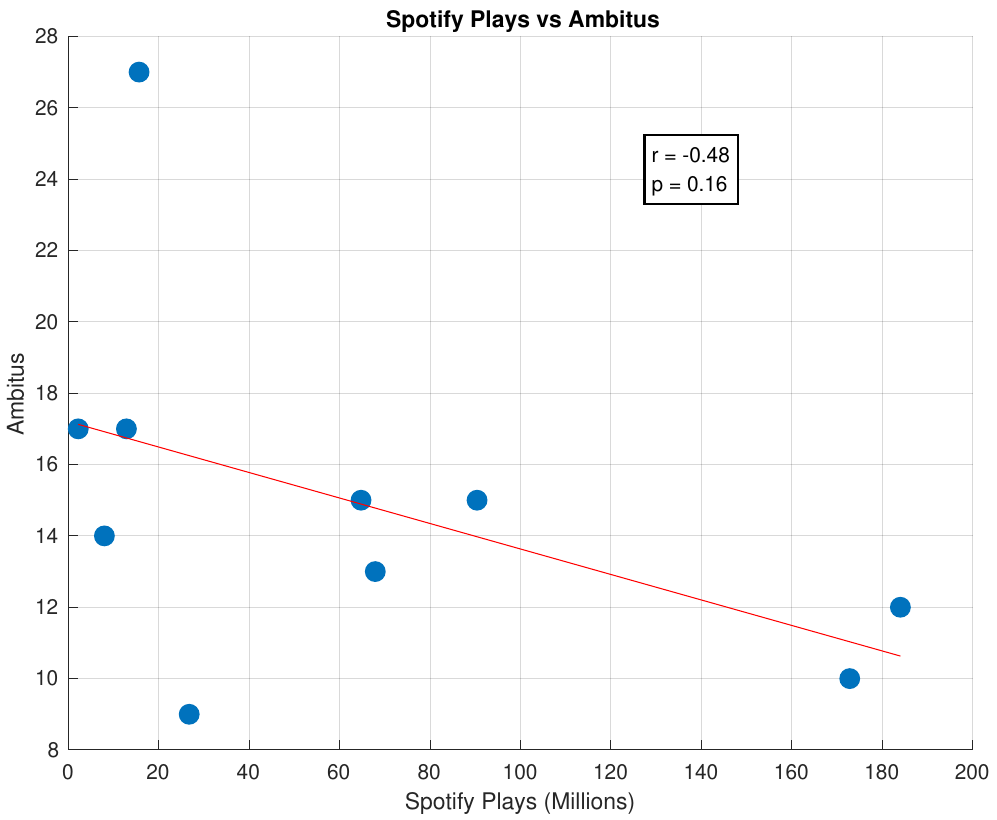}
        \caption{Correlation Between Spotify Plays and Ambitus}
        \label{fig:ambitus2}
    \end{minipage}
\end{figure}

The ambitus results for the analyzed Mexican vocal melodies are detailed in Table \ref{tab:ambitus}. The average ambitus across these songs was 14.27 semitones (Figure \ref{fig:ambitus1}).

The correlation between ambitus and Spotify plays is illustrated in Figure \ref{fig:ambitus2}, revealing a Pearson correlation coefficient of approximately -0.476, with a p-value of approximately 0.165.

\begin{table}[htbp]
    \centering
    \begin{minipage}{0.48\textwidth}
        \centering
        \caption{Ambitus of Selected Mexican Vocal Melodies}
        \label{tab:ambitus}
        \begin{tabular}{|l|c|}
        \hline
        \textbf{Song Title} & \textbf{Ambitus (Semitones)} \\ \hline
        Guadalajara & 27 \\ 
        Cielito Lindo & 17 \\ 
        Mexico Lindo y Querido & 17 \\ 
        Besame Mucho & 15 \\ 
        El Rey & 15 \\ 
        La Cucaracha & 14 \\ 
        Somos Novios & 13 \\ 
        Sabor A Mi & 12 \\ 
        La Bamba & 10 \\ 
        El Son De La Negra Low Lead & 9 \\ 
        El Son De La Negra High Lead & 8 \\ 
        \hline
        \end{tabular}
    \end{minipage}
    \hfill
    \begin{minipage}{0.48\textwidth}
        \centering
        \caption{Pitch Class Entropy of Representative Mexican Vocal Melodies}
        \label{tab:pitch_class_entropy}
        \begin{tabular}{|l|r|}
        \hline
        \textbf{Song Title} & \textbf{Entropy} \\ \hline
        Somos Novios & 0.854020 \\
        Guadalajara & 0.778169 \\
        Sabor A Mi & 0.741163 \\
        El Rey & 0.737659 \\
        La Cucaracha & 0.724993 \\
        Mexico Lindo y Querido & 0.722081 \\
        Besame Mucho & 0.717530 \\
        Cielito Lindo & 0.706387 \\
        El Son De La Negra & 0.699415 \\
        La Bamba & 0.682210 \\
        \hline
        \end{tabular}
    \end{minipage}
\end{table}

\subsection{Presentation of Results}
\subsubsection{Entropy}
\begin{figure}[htbp]
    \centering
    \begin{minipage}{0.48\textwidth}
        \centering
        \includegraphics[width=\linewidth]{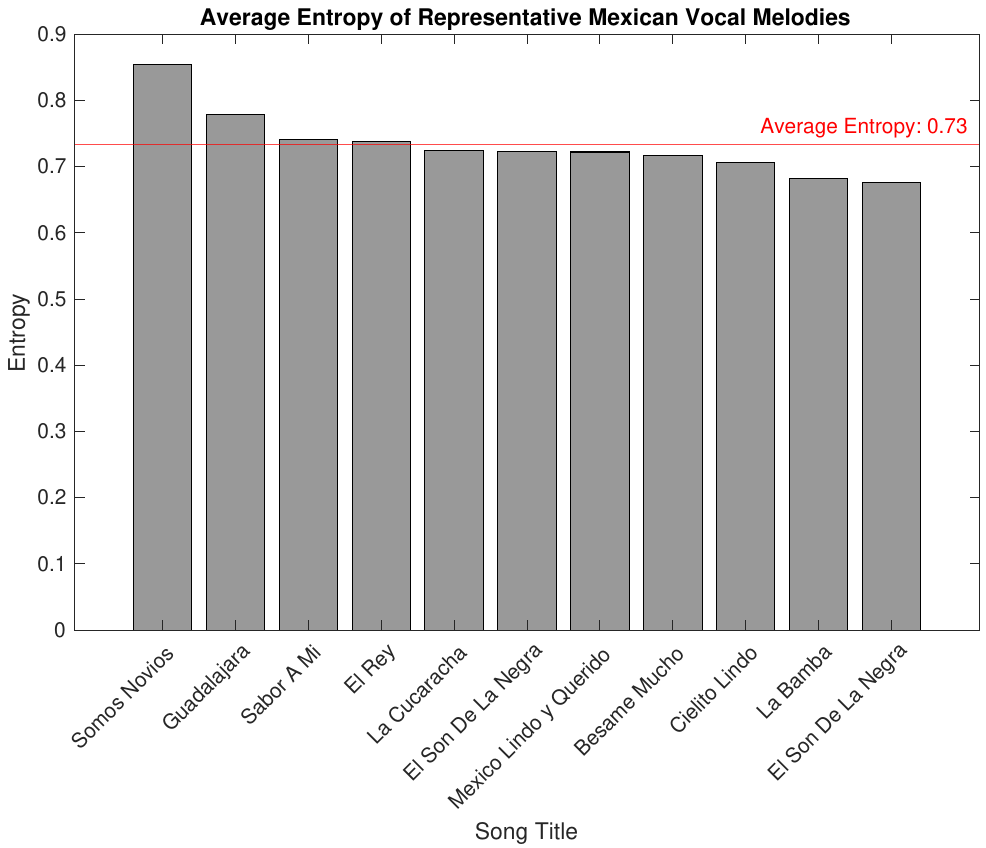}
        \caption{Average Pitch Class Entropy of Representative Mexican Vocal Melodies}
        \label{fig:entropy_avg}
    \end{minipage}
    \hfill
    \begin{minipage}{0.48\textwidth}
        \centering
        \includegraphics[width=\linewidth]{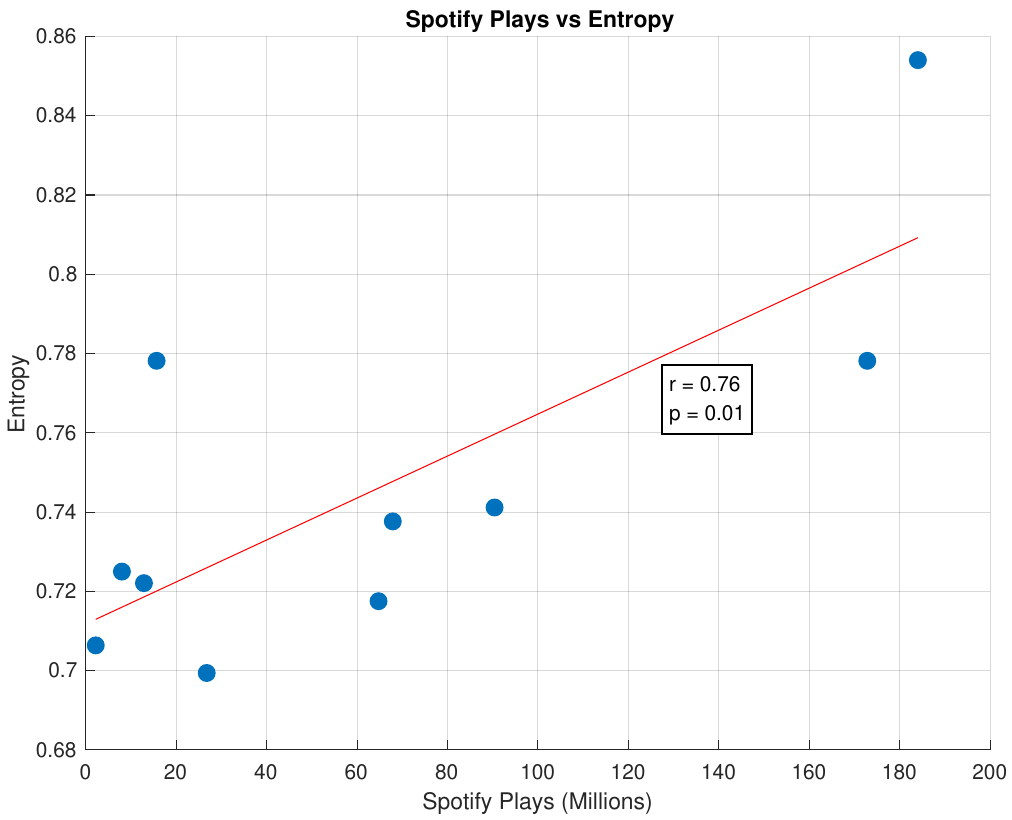}
        \caption{Correlation Between Spotify Plays and Pitch-Class Entropy}
        \label{fig:entropy_corr}
    \end{minipage}
\end{figure}

The pitch class entropy results for the selected Mexican vocal melodies are presented in Table \ref{tab:pitch_class_entropy}. The average entropy of these melodies is 0.73 (Figure \ref{fig:entropy_avg}).

The correlation between entropy and Spotify plays is shown in Figure \ref{fig:entropy_corr}, yielding a Pearson correlation coefficient of approximately -0.0775 with a p-value of approximately 0.843.

\subsubsection{Interval Distribution}
\begin{figure}[htbp]
\centering
\includegraphics[width=0.5\textwidth]{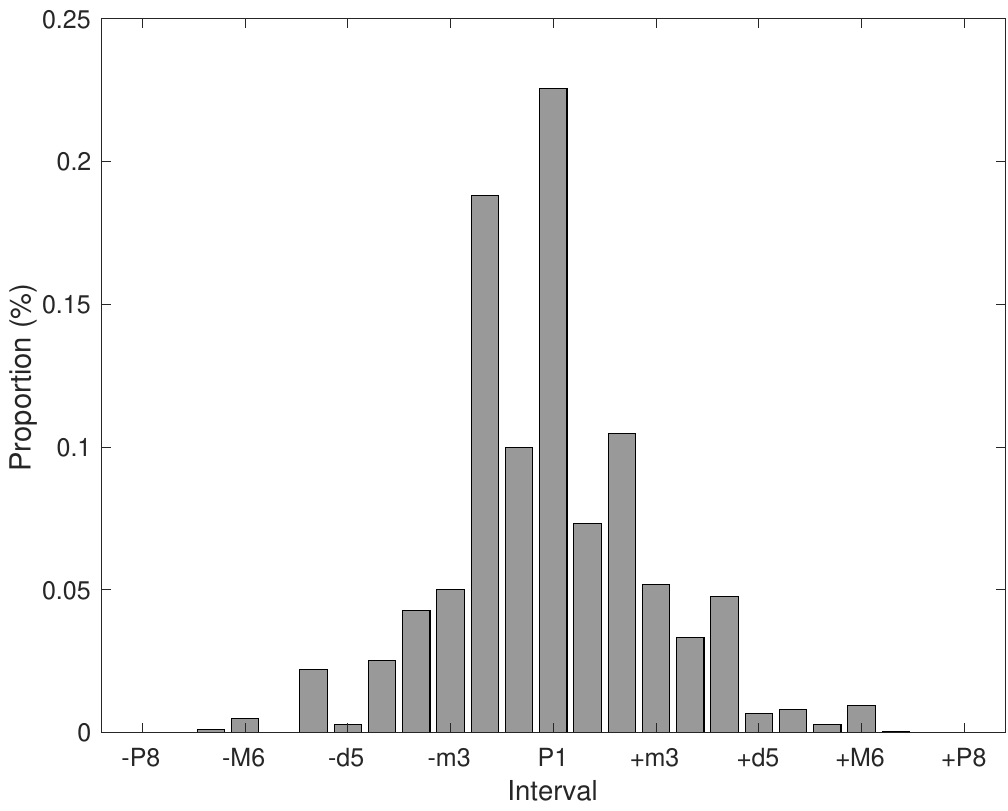}
\caption{Interval Distribution of Representative Mexican Vocal Melodies}
\label{fig:Interval Distribution of Representative Mexican Vocal Melodies}
\end{figure}

As seen in Figure \ref{fig:Interval Distribution of Representative Mexican Vocal Melodies}, the interval distribution graph for the representative Mexican vocal melodies dataset indicates that the prime interval (P1) is the most common interval. This is followed by the major second (M2) and minor second (m2) intervals. The major third(M3) and perfect fourth (P4) are also present but to a lesser extent. The remaining intervals are observed with significantly lower frequencies compared to the more prevalent intervals mentioned earlier.

\subsection{Interpretation of Results}
The ambitus analysis of Mexican vocal folk melodies showcases a remarkable range of compositional styles and vocal demands. At one end of the spectrum, songs like `Guadalajara' have an ambitus of 27 semitones, suggesting a need for advanced vocal skills similar to those required in operatic singing. This implies that some Mexican folk melodies are compositionally complex, requiring skilled vocalists for their expressive execution. On the other end, pieces like `El Son De La Negra High Lead', with an ambitus of just eight semitones, demonstrate the genre's inclusivity, accommodating a broader array of vocal abilities. This contrast within the genre exemplifies Mexican folk music's rich diversity, encompassing intricate compositions for trained singers and more straightforward melodies accessible to a broader audience. Such diversity not only underlines the multifaceted nature of the genre but also its appeal to a range of vocal proficiencies.

In examining the relationship between the ambitus of vocal melodies in representative Mexican songs and their Spotify plays, the statistical analysis revealed a moderate negative Pearson correlation coefficient of approximately -0.476. This coefficient suggests a slight tendency for songs with a larger ambitus to receive fewer plays on Spotify. However, it is crucial to understand the significance of the p-value in this context, which was found to be around 0.165.

A p-value measures the probability that an observed difference could have occurred just by random chance. The standard threshold for statistical significance is a p-value of 0.05 or less. The p-value exceeds this threshold, which implies that the correlation observed might not be statistically meaningful and could be due to random variation in the data. Therefore, despite the initial indication of a correlation between larger ambitus and fewer Spotify plays, the lack of statistical significance (as shown by the p-value higher than 0.05) means a meaningful link between these two variables cannot be confidently asserted.

In the entropy analysis of Mexican vocal melodies, the composition of Armando Manzanero, `Somos Novios' displayed notably high entropy, ranking first. This observation is consistent with Manzanero's status as one of the most prominent and acclaimed Latin music composers\citeyearpar{grammy-manzanero-obituary}. His sophisticated approach to melody is evident in this song's varied and complex nature. In contrast, `La Bamba' rooted in traditional Mexican folk music, showed the lowest entropy, highlighting its simpler, more repetitive structure. This range in entropy values from `La Bamba' to Manzanero's work illustrates the diverse melodic complexity within Mexican music.

The Pearson correlation coefficient between Spotify plays and entropy for the selected songs is approximately -0.0775. This value indicates a weak negative correlation, suggesting no significant linear relationship between the entropy of a song's vocal melody and its number of Spotify plays.

The p-value associated with this correlation is approximately 0.843, which is well above the expected threshold of 0.05 for statistical significance. This high p-value indicates that the weak correlation observed is likely due to random chance rather than a meaningful relationship.

In summary, the statistical analysis does not support a relevant connection between the entropy of vocal melodies in the selected Mexican songs and their popularity on Spotify, as measured by the number of plays.

The interval distribution analysis of the selected Mexican vocal melodies provides valuable insights into their compositional makeup. The prevalence of the prime interval (P1) and major second (M2) in these songs suggests a composition style that favours close, contiguous intervals. This pattern contributes to creating intimate and approachable melodies, resonating with listeners through their simplicity and directness. The inclusion of minor second (m2) intervals introduces a subtle complexity, adding depth to the melodies without overwhelming their fundamental character. The occasional use of major third (M3) and perfect fourth (P4) intervals, though less common, enhances the melodic diversity within each piece. This nuanced approach to interval choice reflects a thoughtful balance between simplicity and musical richness, characteristic of the songs in this analysis. The overall tendency towards smaller intervals in these Mexican vocal melodies highlights their unique charm and accessibility, making them appealing and relatable to a broad audience.

\section{Conclusion}
The exploration of Mexican vocal melodies through MIDI feature extraction has provided a comprehensive view of the structural nuances of this music genre. With a focus on ambitus, pitch-class entropy and interval distribution, the study aimed to decipher the characteristics that define these melodies and their implications for understanding Mexican folk music.

The ambitus analysis revealed a wide range in the melodic scope of the selected songs, indicative of the genre's diverse vocal demands. The ambitus of `Guadalajara' stood out as exceptionally expansive, hinting at a piece that requires significant vocal prowess, likely catering to performers with advanced technical skills. Conversely, melodies with a narrower ambitus, like `El Son De La Negra', may be more accessible for communal singing and enjoyment. When examining the relationship between ambitus and Spotify plays, the data suggested a moderate negative correlation, indicating that songs with a larger ambitus do not necessarily garner more plays. This insight challenges the notion that melodic complexity, as represented by ambitus, correlates with popularity on digital music platforms.

Entropy, which measures the unpredictability of pitch class distribution, offered a quantifiable perspective on the complexity of these melodies. Although the entropy values did not significantly correlate with Spotify plays, the entropy analysis in isolation provided an objective measure of the compositional intricacy of Mexican folk songs. Higher entropy values, like those observed in `Somos Novios', indicate more complex and less predictable melodies, while lower values, such as in `La Bamba', suggest a more repetitive and potentially more accessible melodic structure.

The interval distribution was also a focal point of this study. By analyzing the frequency and duration of intervals within the melodies, we gained insight into the familiar patterns contributing to Mexican folk music's distinctive sound. For example, the prevalence of certain intervals, such as the unison, major seconds and minor seconds, may reflect traditional harmonic and melodic practices unique to this musical culture.

In conclusion, the findings from the ambitus, entropy, and interval distribution analyses have provided valuable insights into the inherent properties of Mexican vocal melodies. While the statistical analysis of Spotify plays did not show a strong correlation with ambitus or entropy, the study has highlighted that the appeal of these songs extends beyond the quantifiable aspects of their structure. Cultural significance, lyrical content, and historical context are among the myriad factors influencing listener preferences and music's digital consumption. This research has laid the groundwork for future studies, which could expand the scope of analysis to include a broader dataset and incorporate qualitative methodologies to capture the whole essence of musical enjoyment and appreciation.

Reflecting on the broader implications of our findings, it becomes evident that the representation of female authorship among the selected songs is notably scarce, with only one song originating from a female artist. This observation points to a larger trend within the music industry, where gender disparities in recognition and representation persist. Additionally, a closer examination of the lyrical content of some songs reveals that, despite their cultural significance, they may perpetuate traditional gender roles or contain elements that could be viewed as aggressive towards women. These insights do not detract from the cultural value of these songs but rather highlight areas for further scholarly inquiry and discussion.

\bibliographystyle{unsrtnat}
\bibliography{references}  






\end{document}